\documentclass[a4paper,fleqn,usenatbib]{mnras}

\usepackage{graphicx}	
\usepackage{amsmath}	
\usepackage{amssymb}	
\usepackage{multicol}        
\usepackage{bm}		
\usepackage{pdflscape}	

\usepackage[T1]{fontenc}
\usepackage{ae,aecompl}

\usepackage{txfonts}

\usepackage{etoolbox}

\makeatletter
\patchcmd\@combinedblfloats{\box\@outputbox}{\unvbox\@outputbox}{}{\errmessage{\noexpand patch failed}}
\makeatother

\title[Multiplicity Matters]{Accounting for Multiplicity in Calculating Eta Earth} 

\author[Zink and Hansen]{Jon K. Zink$^{1,2}$ \thanks{E-mail: \href{mailto:jzink@astro.ucla.edu}{jzink@astro.ucla.edu}}, and Bradley M. S. Hansen$^{1}$
\\
$^{1}$Mani L. Bhaumik Institute for Theoretical Physics, Department of Physics and Astronomy, University of California, Los Angeles, CA 90095\\
$^{2}$Caltech/IPAC-NASA Exoplanet Science Institute, Pasadena, CA 91125}

\date{Last updated 2019 May 2}

\pubyear{2019}

\begin{document}
\pagerange{1--7}
\maketitle

\begin{abstract}
Using the updated exoplanet population parameters of our previous study, which includes the planetary radius updates from \emph{Gaia} DR2 and an inferred multiplicity distribution, we provide a revised $\eta_{\earth}$ calculation. This is achieved by sampling planets from our derived population model and determining which planets meet our criterion for habitability. To ensure robust results, we provide probabilities calculated over a range of upper radius limits. Our most optimistic criterion for habitability provides an $\eta_{\earth}$ value of $0.34\pm 0.02 \frac{\rm planets}{\rm star}$. We also consider the effects of multiplicity and the number of habitable planets each system may contain. Our calculation indicates that $6.4\pm0.5\%$ of GK dwarfs have more than one planet within their habitable zone. This optimistic habitability criterion also suggests that $0.036\pm0.009\%$ of solar-like stars will harbor 5 or more habitable planets. These tightly packed highly habitable system should be extremely rare, but still possible. Even with our most pessimistic criterion we still expect that $1.8\pm0.2\%$ of solar-like stars harbor more than one habitable planet.

\end{abstract}


\section{Introduction} 
As the number of known exoplanets continues to grow, understanding the frequency of habitable planets remains a major area of interest. Assuming extraterrestrial life will share our fundamental characteristics, the search for habitability is focused on finding rocky planets orbiting a star at a distance that allows for the existence of liquid water. It is currently unclear which stellar types help or hinder the habitability of a planet. Several studies have considered habitability around M dwarf stars \citep{yan13,kop13b,dre13,dre15,lug15}, but tidal effects and UV radiation make habitability unclear. Knowing that life is possible around solar-like stars, this study focuses on GK dwarf hosts (stars with $4200<T_{\rm eff}<6100K$) and the possibility of retaining habitable planets.\

The \emph{Kepler} space mission set out to explore the parameter space of Earth analogs. In finding $\sim 4,500$ transiting exoplanets candidates, the continuous viewing of a fixed field makes \emph{Kepler} an ideal data set for occurrence measurements. One of the mission goals was to establish an estimate for $\eta_{\earth}$, which is defined as the expected occurrence of ``Earth-sized'' planets within the habitable zone. However, significant discrepancies exist among the studies that have sought to produce such an estimate: \citet{cat11} found an $\eta_{\earth}$ between 0.01-0.03, \citet{pet13} found $\eta_{\earth}=0.22$ , \citet{sil15} found $\eta_{\earth}=0.064$, \citet{tru16} found $\eta_{\earth}=0.75$, and a recent study by \citet{bar18} found an $\eta_{\earth}$ value of 0.35. One reason for this significant variance is that ``Earth-sized'' is defined as a terrestrial planet that could potentially harbor life. Lacking the knowledge of each planet's composition, planets within $0.5-2~r_{\earth}$ have been considered for this designation. Each of the previous studies have chosen unique limits within this range, creating significant variations in the inferred values. This study emphasizes clear radius bounds and provided several different limits for comparison. Additionally, the complete \emph{Kepler} DR25 is now available \citep{mat17} along with improved planetary and stellar parameters from \emph{Gaia} DR2 \citep{gai18,ber18} and asteroseismology \citep{vanE18a}, enabling updated occurrence measurements of $\eta_{\earth}$ to be performed.\

Our previous study \citep{zin18} found that detection efficiency is a function of multiplicity. Using the injection recovery results of \citet{chr17}, we determined that the first detected TCE (Transit-Crossing Event) within each systems was recovered with a higher efficiency then any subsequent TCEs within a given light curve. In other words, when using the \emph{Kepler} pipeline the first detected planet will be easier to detect than another other additional planets within the system. From this result it is apparent that the population of single planet systems is over represented in the empirical \emph{Kepler} data set. It is likely that additional planets exits within the light curves, increasing the expected overall planet occurrence.  Building on this conclusion, we used the updated stellar radius measurements \citep{ber18}, inferred from \emph{Gaia} parallax, to determine current occurrence parameters for the planet radius and period power-laws. With a better understanding of the completeness of multi-planet systems, we calculated a multiplicity distribution for the \emph{Kepler} candidates. Using our updated occurrence estimates --which accounts for the significant loss of multi-planet systems within the \emph{Kepler} data set-- it is now possible to consider system architecture when calculating the probability of a true Earth analog. 

Additionally, the recent discovery of TRAPPIST-1 \citep{gil16}, with up to four planets orbiting in the systems habitable zone, has generated interest in the concept of multiple habitable planets within a single system. The proximity to this M dwarf star (all are planets orbit at <0.06 AU) imposes significant barriers for the planets to retain true habitable status \citep{gar17,pea19}, but it remains unclear if these effects will render the entirety of the known planets uninhabitable \citep{sah18,pap18,dob19,fra19}. Nevertheless, this system of rich multiplicity, provides a unique opportunity to study the differences in atmospherical conditions and the corresponding effects on habitability \citep{lin18}. With the improved multiplicity measurements from \citet{zin18}, it is now feasible to estimate the likelihood of multiple habitable planet systems.\

Our goal in this paper is to provide an updated $\eta_{\earth}$ value and discuss how this value is affected by multiplicity within the habitable zone. In Section \ref{sec:stell} we describe our method of stellar selection for this calculation. In Section \ref{sec:draw} we provide a method of planetary sampling to ensure we minimize orbital instabilities. In Section \ref{sec:habit} we discuss the criterion we deem necessary for a planet to be considered habitable. In Section \ref{sec:results} we present the results of our occurrence calculation and discuss probability of multiple planets within the habitable parameter space. In Section \ref{sec:disc} we discuss the limitations of our calculation. In Section \ref{sec:con} we provide our concluding remarks on this paper.

\section{Stellar Selection}
\label{sec:stell}
Because we are implementing the results of \citet{zin18}, we use the same stellar selection methods. We shall briefly describe this method here, but suggest that the reader consult \citet{zin18} for a more thorough description.  The complete stellar sample is provided by the \emph{Kepler} DR25 stellar parameters \citep{mat17} with correction from \emph{Gaia} DR2 \citep{ber18}. We make various cuts in the sample to ensure the occurrence measurement reflects planets around solar-like GK dwarfs. Our stellar sample includes stars with $4200<T_{\rm eff}<6100K$, $log(g)\ge4$, $R_{\star}\le 2R_{\sun}$, and available mass measurements for each star. Despite making these stellar cuts, we acknowledge that it is still possible for sub-giants to be contaminating our sample. The most extreme example of possible contamination in our sample is KIC 4273493, with $log(g)=4.007$, $R_{\star}=1.991R_{\sun}$, $T_{\rm eff}=5426K$, and $M_{\star}=0.871M_{\sun}$. Given the stellar parameters of this edge case, we do not expect significant sub-giant contamination. To avoid under sampled systems we only include stars with a data span $>2$ years with $>60\%$ coverage during that period. To avoid contamination from especially noisy sources, we exclude stars which exceed 1000 ppm in the $\rm CDPP_{7.5h}$. CDPP (Combined Differential Photometric Precision) is a measure of the amount of noise any planet, transiting for a given period of time, must overcome to provide a $1\sigma$ detection \citep{chr12}. After making the described cuts, we find that 86,605 solar-like stars remain. A machine-readable version of this data is available \href{https://github.com/jonzink/ExoMult}{online}.

\section{Method of Sampling Planets}
\label{sec:draw}
To determine the probability of habitable planets existing, we draw from the power-law and multiplicity distributions found in \citet{zin18} for each star in our stellar sample. \

The multiplicity of the system is drawn from a modified Poisson distribution:

\begin{equation}
p(m)=\kappa\Big(\frac{\lambda^{m}e^{-\lambda}}{(m)!}-e^{-\lambda}\Big) \label{equ:cdfp}
\end{equation}
  
where $m$ in the multiplicity of the system. Using the $\lambda=8.40 \pm 0.31$ and $\kappa=0.70 \pm 0.01$ values found in our previous study, we randomly assign the number of planets each system will contain. The $\kappa$ value represents the number of stars with planets, indicating that about 30\% of our sampled systems will not harbor any planets at all. The $\lambda$ value indicates the average number of planets each planet-harboring system should contain.\

To determine the radius and period of each of these planets we draw from two independent broken power-laws ($g(p)$ and $q(r)$):

\begin{equation}
\label{eq:plaw}	
g(p)= 
\begin{cases}
C_{p1}p^{\beta_1} & \text{if $p<p_{br}$} \\
C_{p2}p^{\beta_2} & \text{if $p \geq p_{br}$} \\
\end{cases}
\end{equation}

\begin{equation}
q(r)= 
\begin{cases}
C_{r1}r^{\alpha_1} & \text{if $r<r_{br}$} \\
C_{r2}r^{\alpha_2} & \text{if $r \geq r_{br}$} \\
\end{cases}
\end{equation}

where $\beta_1$ and $\beta_2$ indicate the slope of the period power-laws and $\alpha_1$ and $\alpha_2$ indicate the slope of the radius power-laws. $r_{br}$ and $p_{br}$ correspond to the breaks in the radius and period distributions receptively. Here, $C_{r1},C_{r2},C_{p1},$ and $C_{p2}$ are normalization constants which force continuity at each of the corresponding breaks. From \citet{zin18} we use $\alpha_1=-1.65\pm^{0.05}_{0.06}$, $\alpha_2=-4.35\pm0.12$, $\beta_1=0.76\pm0.05$, and $\beta_2=-0.64\pm0.02$ for the power-laws and $p_{br}=7.08\pm^{0.32}_{0.31}$ days and $r_{br}=2.66\pm0.06 r_{\earth}$ for the corresponding breaks. The previously described distributions have been fit using a period range of $.5\le p \le 500$ days and a radius range of $.5 \le r \le 16 r_{\earth}$, thus we shall adopt the same sampling range here. \

Our previous study assumed perfect radius measurements for each planet when inferring the mentioned power-laws. Careful consideration is needed when making this assumption, particularly when our initial planet sample had an average relative error of $14\%$ for planet radius. If we assume uniform and symmetric Gaussian noise, the uncertainty in the power-law parameters will account for much of the randomness cause by this mis-measure. However, this is not the case, as many factors determine the uncertainty of radius, creating unique uncertainty values for each planet. Furthermore, if an additional systematic offset is found, similar to the overall increase in stellar radii discovered with \emph{Gaia} DR2 \citep{ber18} (which has been accounted for in this study), our results will significantly differ. Accounting for such biases would require a more detailed fitting of the power-law model and is beyond the scope of this study. In this study, we assume the planet radii are known perfectly.\

Once each planetary system has been drawn, we determine the stability of the system. With the independent draw of period and radius, most system will be overly crowded and lack the necessary spacing required for stability. To quantitatively understand this, we calculate the planet separations ($\Delta H$) in units of mutual Hill radii \citep{cha96} as follows:

\begin{equation}
\Delta H =\frac{a_2-a_1}{(\frac{M_1+M_2}{3*M_{\star}})^{1/3} * \frac{a_1+a_2}{2}}
\end{equation}

where $a_1$ and $a_2$ are the semi major axis of the inner and outer planet orbits respectively, and $M_1$, $M_2$, and $M_{\star}$ correspond to the masses of the inner, outer, and stellar masses respectively. Since our sampling distribution provides a radius value, we must convert the planet radius into mass. This is done by adopting the empirical mass-radius relationship found by \citet{chen17}:

\begin{equation}
M=0.972*\Big(\frac{r}{r_{\earth}}\Big)^{3.584} M_{\earth} \quad\quad ;\quad\quad   r<1.23r_{\earth}\\
\end{equation}

\begin{equation}
M=1.436*\Big(\frac{r}{r_{\earth}}\Big)^{1.698} M_{\earth} \quad\quad ;\quad\quad 1.23r_{\earth}\le r<14.31r_{\earth} \\
\end{equation}

\begin{equation}
M=131.581 M_{\earth} \quad\quad ;\quad\quad  r\ge 14.31 r_{\earth} \\
\end{equation}

The empirical data set contains a significant dispersion in this relationship, making it apparent that planets of similar mass can correspond to different radii. We ignore this issue and assume a perfect correlation between mass and radius. This assumption is justified by the fact that $\Delta H$ scales as $M^{-1/3}$, making small deviations in $M$ insignificant. For example, if we increase the mass of a planet in our sample by $50\%$, the average correction factor needed for $\Delta H$ is a $5\%$ decrease. Additionally, \citet{chen17} finds a weak negative power law for $r>14.13 r_{\earth}$ ($r \propto M^{-0.044}$), while we assume $M=131.581 M_{\earth}$ in this range of r. This simplification is necessary in order for the inverse equation ($M(r)$) to remain a function over the parameters space of interest. \

\begin {table*}
\begin{flushleft}
\caption {The probability that a solar-like star will process the indicated number of habitable planets. The upper radius limit was varied to account for the lack of a clear cut off for rocky planets and therefore habitability. Each row indicates the probability of finding at least the indicated multiplicity of planets within the habitable parameters of this study. The blank cells (-) indicate multiplicities so rare that their occurrence provided no statistical significance (probability$<3\sigma$). The $\eta_{\earth}$ values correspond to the sum of all multiplicities. All $\eta_{\earth}$ values are given in units of $\frac{\rm planets}{\rm star}$. \label{tab:habTable}} 
\begin{tabular*}
{\textwidth}{@{\extracolsep{\fill}}l c c c c c c  } 
\hline \hline \multicolumn{1}{l}{} & \multicolumn{1}{c}{$0.72\le r\le 1.00 r_{\earth}$} & \multicolumn{1}{c}{$0.72\le r\le 1.23 r_{\earth}$} & \multicolumn{1}{c}{$0.72\le r\le 1.48 r_{\earth}$} & \multicolumn{1}{c}{$0.72\le r\le 1.62 r_{\earth}$} & \multicolumn{1}{c}{$0.72\le r\le 1.70 r_{\earth}$}\\
\hline 
            One &   $0.139\pm0.007$ &   $0.198\pm0.007$ &    $0.237\pm0.008$ &   $0.253\pm0.009$  &   $0.261\pm0.009$\\
            Two &   $0.018\pm0.002$ &   $0.036\pm0.003$ &   $0.052\pm0.004$ &   $0.060\pm0.005$ &   $0.064\pm0.005$\\
          Three & $0.0016\pm0.0002$ & $0.0049\pm0.0007$ & $0.0088\pm0.0013$ &   $0.011\pm0.002$ &   $0.012\pm0.002$\\
           Four &                 - & $0.00058\pm0.00013$ & $0.0014\pm0.0003$ & $0.0020\pm0.0004$ &   $0.0023\pm0.0004$\\
           Five &                 - &                 - &       $0.00019\pm0.00006$ & $0.00029\pm0.00008$ &   $0.00036\pm0.00009$\\
\hline
$\eta_{\earth}$ &   $0.158\pm0.008$ &     $0.24\pm0.01$ &      $0.30\pm0.01$ &      $0.33\pm.02$   &     $0.34\pm.02$ \\
\hline

\end{tabular*}
\end{flushleft}
\end{table*}

Having provided a method for calculating mass from radius, we can now calculate the $\Delta H$ of each planet spacing. Upon our initial draw we find that roughly 50\% of all spacings fall above $\Delta H=10$. Empirically it has been shown that 93\% of \emph{Kepler} multi-planet spacings have a $\Delta H\ge10$ \citep{fan12,pu15,wei17}. This threshold indicates a spacing large enough to avoid instabilities cause by two planets in opposition. It is worth noting that orbits with larger eccentricity will require even greater spacing to avoid such interactions \citep{daw16}. However, most models for the formation of these systems favor low eccentricities. In situ assembly models yield $\langle e \rangle =0.11$ \citep{han13}, while chain migration models require small eccentricities to maintain the stability of the chain \citep{gol14}. For this study we ignore such issues and only consider planets with circular orbits. To avoid testing unstable systems, we resample all systems with any spacings of $\Delta H <10$. In this resample we maintain the initial multiplicity (to avoid the artificial exclusion of high multiplicity systems), but re-draw all period and radius measurement within the system, regardless of the spacing that failed to meet this requirement. This type of complete resampling ensures that we maintain an independent broken power-law distributions of both period and radius. The $\Delta H$ values are then re-calculated and if all spacing now exceeds or meets the $\Delta H=10$ requirement, the system is no longer re-drawn.  However, if any spacing in the system remains below this threshold, the planets are again re-drawn. This process will continue until either all spaces meet the $\Delta H$ requirement or 100 iterations have occurred. We stop after 100 iterations to mimic the $\Delta H <10$ spacings that exist in the empirical data set. In testing, we found that systems with more than 10 planets almost always fail to converge to the $\Delta H <10$ spacing requirement even after 100 iterations. To avoid these extremely unstable planetary archetypes, we allow a maximum of 10 planets per system. After this procedure has been completed we find that roughly 90\% of the spacings now meet or exceed this $\Delta H$ threshold (although this value can be as low as 85\% and as high as 95\%). When compared to the \emph{Kepler} sample (93\% of planet separations have $\Delta H\ge 10$), we produce a slightly smaller, but still within statistical variability, average spacing. \

For clarity, we provide here the important statistics of a given run of our simulation. 26,057 stars have no planets and 60,548 stars have at least one planet. The number of systems with 1:10 planets respectively: 692, 1828, 3619, 5731, 7565, 8559, 8473, 7455, 5904, 10722. The apparent pile up of 10 planet systems is cause by our multiplicity distribution, which anticipates a non-negligible number of systems with more than 10 planets. When we remove planets from these systems (to meet our previously described 10 planet maximum), we create a mild surplus of 10 planet systems. After 100 iterations of drawing systems, 91\% of the planet separations are of $\Delta H\ge 10$ with 90\% of the total planets sample having a $r<2r_{\earth}$ and 44\% of the total planet sample with a $p>300$days.\

\section{Habitability Requirements}
\label{sec:habit}	

In order to determine which planets could be habitable in our simulation we must impose some habitability criterion. We shall focus on two main factors that determine the habitability of a planet: location and size.\

The mass of the planet plays an essential role in habitability. If a planet is too small it will not be able to retain an atmosphere and maintain ongoing plate tectonic activity \citep{kas93}. Since planet cooling rates scale as $r^{-1}$, smaller mass/radius planets will cool much faster. This will lead to hardening of the planet interior, which in effect will significantly decrease the planets magnetic field that shields from cosmic radiation and atmospherical stellar wind stripping \citep{bre03}. Furthermore, the thick crust of the planet will halt any plate tectonic activity \citep{one07,val07,kit09,noa14}. This lack of active geology will not benefit from the long-term climate stabilization produced by the $CO_2$ cycle \citep{kas03,coc16,rus18}. Using the radiogenic flux model provided by \citet{wil97}, \citet{ray07} established that planets with roughly $M<0.3M_{\earth}$ will lack this important geological activity. Therefore, we only consider planets habitable with a $M\ge0.3M_{\earth}$. Using our conversion model in Section \ref{sec:draw}, this mass cut-off corresponds to $r\ge0.72r_{\earth}$. \

The maximum mass limit for habitability is set by the limit in which the planet is no longer rocky, but exists as a gas giant. This limit is difficult to determine as the composition of the planet is really the important factor to considers. Observationally, there exists low mass planets like Kepler-11f ($M=2.3\pm^{2.2}_{1.2}M_{\earth}$) with large radii ($r=2.61\pm0.25r_{\earth}$), indicating a gaseous planet \citep{lis11}. Conversely, the existence of large mass rocky planets like Kepler-10c ($M=7.4\pm^{1.3}_{1.2}M_{\earth}$; \citealt{raj17}) with a radii of $r=2.35\pm^{0.09}_{0.04}r_{\earth}$ have made mass limits difficult to pin down \citep{dum14}. Considering the entirety of the known population, the mass-radius relationship for known exoplanets produces a break around $4M_{\earth}$ \citep{wei14, wol16}. This break indicates a general transition from rocky to gaseous. However, using a interior planet model and Bayesian analysis, \citet{rog15} found a $95\%$ confidence limit for the rocky planet transition at $r=1.62\pm^{0.67}_{0.08}r_{\earth}$, with a best fit transition occurring at $r=1.48\pm^{0.08}_{0.04}r_{\earth}$. Using the empirical mass-radius sample, \citet{chen17} found a similar transition value of $r=1.23^{0.44}_{0.22}r_{\earth}$. A more recent study, which incorporated the \emph{Gaia} DR2 data, found a gap in the planet radius population around $1.7r_{\earth}$, signifying a transition from rocky to gaseous \citep{ful18}. We also know that the Earth is rocky and habitable, therefore providing a pessimistic limit of $1r_{\earth}$.  Because it remains unclear where this transition takes place, we shall provide the results for an upper radius limit of $1r_{\earth}$, $1.23r_{\earth}$, $1.48r_{\earth}$, $1.62r_{\earth}$ and $1.70r_{\earth}$ in Section \ref{sec:results}.\

The location of the planet is important for habitability as it receive enough stellar incident flux to support liquid water. Furthermore, the planet must not be so close to the star as to undergo a runaway greenhouse effect. This region is known as the Habitable Zone (HZ hereafter; \citealt{hua59,kas93,kop13}). Current versions of this region account for the mass (surface gravity) of the planet and how atmospheric $H_2O, CO_2,$ and $N_2$ will affect the planets ability to retain heat \citep{kas93,kop14,ram14}. The effective incident stellar flux ($S_{\rm eff}$) is given by a fourth order polynomial fit of the effective stellar temperature ($T_{\rm eff}$):

\begin{equation}
\label{eq:seff}	
\begin{split}	
S_{\rm eff}(M,T) =S_{\star}(M)+a(M)*T+b(M)*T^2\\
+c(M)*T^3+d(M)*T^4
\end{split}	
\end{equation}
 
where $S_{\star}(M),a(M),b(M),c(M),d(M)$ are all parameters fit the account for the mass of the planet and $T$ is the normalized stellar temperature ($T=T_{\rm eff}-5780K$). We adopt the polynomial values provided by \citet{kop14}, linearly interpolating with mass between the provided ($.1,1,5M_{\earth}$) parameters. Using the ``Runaway Greenhouse limit'' provides a conservative estimate for the inner edge of the HZ. The physical distance ($d$) of this limit can be found by using the $S_{\rm eff}$ value derived in Equation \ref{eq:seff}:

\begin{equation}
d=\sqrt{\frac{L/L_{\sun}}{S_{\rm eff}}} \rm AU
\end{equation}

where $L$ is the luminosity of the host star. The outer HZ limit is defined by the ``Maximum Greenhouse limit'', where $CO_2$ partial pressure remains just high enough to produce any amount of greenhouse heat retention. This border is independent of planet mass and fixed only by the host star's $T_{\rm eff}$. We consider planets, that lay between these two limits, habitable. \

Because the habitable zone is dependent on stellar $T_{\rm eff}$ and many of the values have been estimated through photometry, we consider how fluctuations in these inferred values may affect our population. To do so, we sample from a Gaussian distribution centered around each star's expected $T_{\rm eff}$, using the parameter uncertainty for the distribution width. Within our sample, the median value of $\sigma _{T_{\rm eff}}$ is 180K with a standard deviation of 33K, corresponding to an average relative error of $\sim3\%$ for $T_{\rm eff}$. Using these uncertainties we resample $T_{\rm eff}$ during each iteration of this study. Any inaccuracies in the expected values will be accounted for using this method.  

In summary, we only consider a planet habitable if the semi-major axis of the planet lays within the bounds provided by the \citet{kop14} HZ model (0.95-1.68 AU for an Earth mass planet around a solar-like star) and the planet radius lays between $0.72\le r\le 1.70R_{\earth}$.\  

\section{Results}
\label{sec:results}

Using the updated radius, period, and multiplicity distributions from \citet{zin18}, we randomly assigned planets to the \emph{Kepler} stellar sample. From this sampling we determined which planet are habitable using the criterion discussed in Section \ref{sec:habit}. Taking the number of habitable planets and dividing by the total number of stars in our stellar sample (86,605), provided us with a statistical value for habitability. This process was completed 100 times to account for variations in the distribution models (allowing movement within the uncertainty of these parameters: $T_{\rm eff}$, $\alpha_1$, $\alpha_2$, $\beta_1$, $\beta_2$, $p_{br}$, $r_{br}$, $\lambda$, and $\kappa$) and system archetypes. We calculate the probabilities for each of the 100 runs separately. The standard deviation between each of the 100 runs is used to calculate the uncertainty in our results. In testing we found no significant difference in our results when sampled with multiple runs of 50, indicating that 100 runs provides more than sufficient coverage of the uncertainty parameter space. In Table \ref{tab:habTable} we present the results of this simulation.\      

Using the multiplicity distributions of \citet{zin18} provides an opportunity to understand the occurrence of multiple habitable planets within a given system. In our most optimistic upper radius limit ($0.72r_{\earth}\le r\le1.70r_{\earth}$), we find that multiplicity within the HZ should occur for GK dwarfs with a probability of $0.063\pm0.005$. This limit also provides statistical significance for systems with as many as five habitable planets. These extremely rare systems (probability=$0.00036\pm0.00009$) consist mostly of tightly packed small mass planets that spread across the entirety of the HZ. Within our stellar sample we would expect 31 systems to harbor this unique architecture. These systems tend to have a $\Delta H < 10$ between several of the planets with an average separation of $\overline{\Delta H}$=6.8. When compared with other observed compact systems: Kepler 11 (b,c,d,e,f; $\overline{\Delta H}$=9.4), Kepler 90 (d,e,f,g,h; $\overline{\Delta H}$=8.7), and TRAPPIST-1 (d,e,f,g,h; $\overline{\Delta H}$=9.4), our simulated systems appear to be even more tightly packed. This could indicated that several of our simulated systems with five habitable planets are unstable, decreasing the probability of such systems existence. However, \citet{obe17} found through simulation that five Earth sized planets could survive in the habitable zone of a Sun-like star for at least $10^9$ orbits, providing evidence that stable versions of these systems are possible. \

When we consider our most pessimistic radius limit ($0.72r_{\earth}\le r \le 1r_{\earth}$), we still anticipate the existence of multiple habitable planet systems (probability=$0.018\pm0.002$) with as many as three habitable planets within a single system (probability=$0.0016\pm0.0002$). These system are all very stable with almost all spacings exhibiting a $\Delta H$>10 in our simulation. Although more rare, we should still expect to find multiplicity within the HZ for a significant fraction of GK dwarf systems. The intermediate upper radius limits are provided to show how these values change as we transitions from the lower limit of Earth ($0.72\le r\le1.00r_{\earth}$) to the upper limit of \citet{ful18} ($0.72\le r\le1.70r_{\earth}$).\   

The $\eta_{\earth}$ value corresponds the frequency in which you would expect to find a habitable planet. Since all previously studies have marginalized over multiplicity, we use this value for comparison. Using our updated population parameters we find an optimistic value of $\eta_{\earth}=0.34\pm 0.02 \frac{\rm planets}{\rm star}$ for solar-like stars. This is consistent with \citet{tru11}, who found $\eta_{\earth}=0.34\pm 0.14$ using the first 136 days of \emph{Kepler} data with a habitable radius range of $0.5\le r \le 2.0 r_{\earth}$ and HZ of 0.8-1.8 AU. Using the same 134 day data set, \citet{cat11} found a range of 0.01-0.03 for $\eta_{\earth}$ with a more restricted habitable radius of $0.8\le r \le 2.0 r_{\earth}$ and a HZ of 0.75-1.8 AU. This discrepancy was caused by a lack of accounting for completeness and the inclusion of planets beyond 42 day periods. A more complete \emph{Kepler} sample (Q1-16) study came from \citet{pet13}, who found $\eta_{\earth}=0.22\pm 0.08$ implementing a 0.25-4 solar flux limit (0.5-2 AU for a solar analog) and 1-2$r_{\earth}$ radius limit for habitability. \citet{pet13} only used the largest SNR planet from each system, avoiding any issues with multiplicity. For proper comparison we should consider our optimistic model with only one habitable planet ($0.261\pm 0.009$), which is within the uncertainty of \citet{pet13}. Again using the same data set and radius restrictions as \citet{pet13} but with a much more narrow HZ range (0.99-1.7 AU), \citet{sil15} found a smaller probability ($0.064\pm^{0.034}_{0.011}$). Using a forward model method, \citet{tru16} found an $\eta_{\earth}$ value of $0.75\pm 0.11$ and $1.03\pm 0.10$ for K and G type stars respectively. Here, the range of 0.8-1.8 AU and 0.5-1.25 $r_{\earth}$ was considered habitable, inflating the calculated values. Additionally, the separation of planets by stellar spectral type reduces the number planets used for each extrapolation, producing significant deviations from the previously calculated values. Various assumptions about the habitable radius range and HZ cause much of the dispersion seen among these values. Recently, \citet{bar18} used a simple population extrapolation with \emph{Gaia} updates and found $\eta_{\earth}=0.35\pm^{0.25}_{0.16}$ for a radius range of 1-1.75 $r_{\earth}$ and a flux limit of 0.2-2 solar flux (0.7-2.2 AU for a solar analog). This simple method produced a value comparable to our more detailed calculation.  Our clear habitability assumptions and updated distributions parameters provide the most current and complete $\eta_{\earth}$ value. \   

Within this study we use conservative HZ limits (0.95-1.68 AU for an Earth mass planet around a solar-like star) corresponding to the location of runaway and maximum greenhouse effect. However, if we consider a more optimistic range using the flux received by recent Venus (0.75 AU) and early Mars (1.78 AU) \citep{kop13}, we find significantly larger $\eta_{\earth}$ values. For our upper radius limits of 1,1.23,1.48,1.62, and 1.7$r_{\earth}$, we find $\eta_{\earth}$ is equal to $0.27\pm0.01$,$0.39\pm0.02$ ,$0.48\pm0.02$ ,$0.52\pm0.02$, and $0.55\pm0.02$ respectively. This is nearly a 60\% increase at each radius limit. However, these optimistic limits are inspired by assumptions of past conditions on Venus and Mars, and ignore the effects of planet mass on the HZ, providing an upper limit on the HZ range \citep{kas93}.

\begin{figure}
\begin{center}
\includegraphics[width=8.5cm]{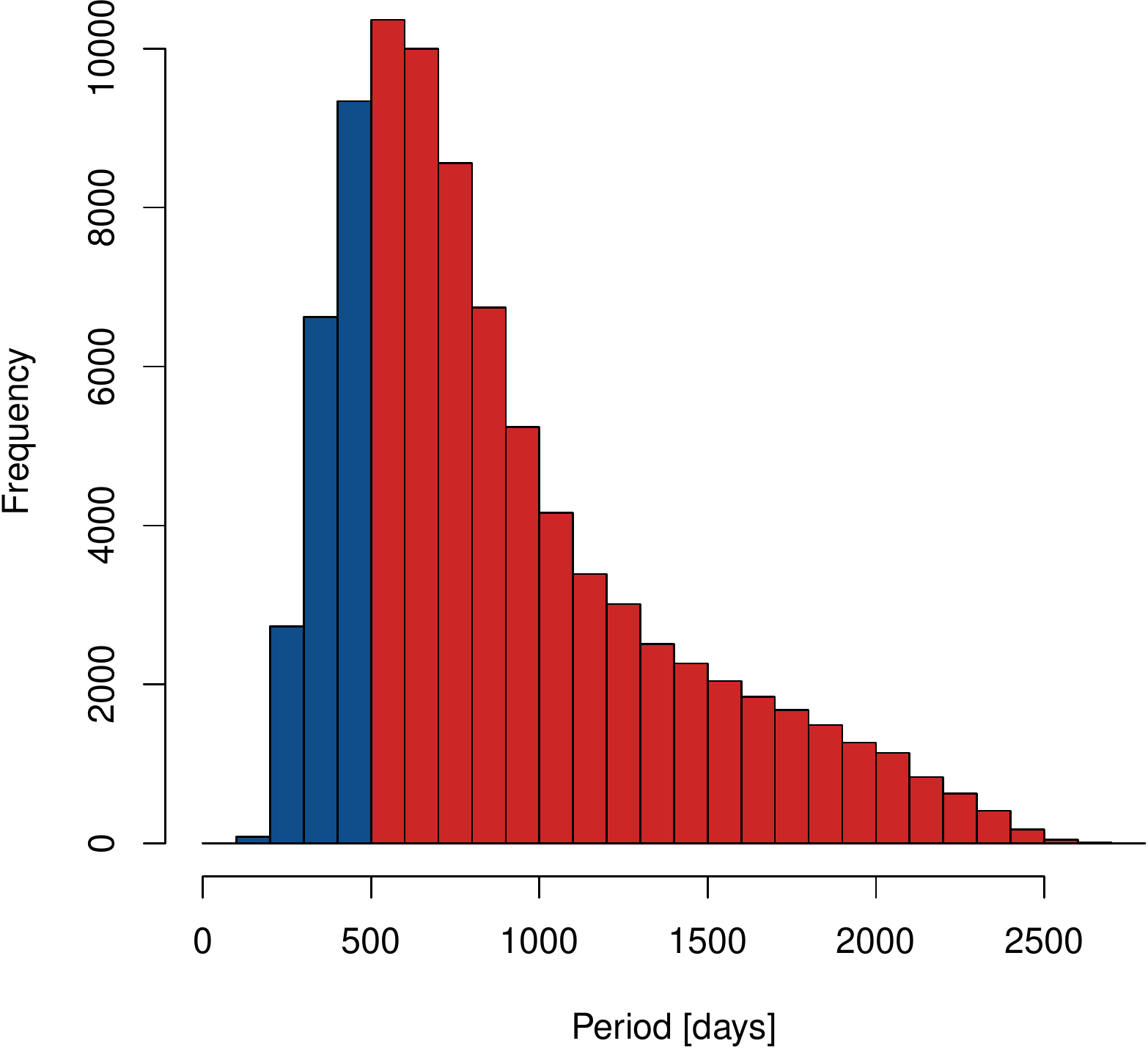}
\end{center}
\caption{A histogram showing the outer HZ limit for the stellar sample of this study. The blue corresponds to the HZ limits within the 500 period range of our calculation and the red corresponds to the stars that have a HZ extending beyond a 500 day period. \label{fig:hzcut}}
\end{figure}

\section{Discussion}
\label{sec:disc}
The radius distribution inferred from the \emph{Kepler} data set provides sufficient coverage of the important radius range for habitability, extending from $0.5r_{\earth}\le r \le 16r_{\earth}$. This is well beyond the limits of $0.72r_{\earth}\le r \le 1.62r_{\earth}$ that we considered habitable. However, it is important to remember that almost all of the known exoplanets of this size have short periods, due to the low signal produced by such small radii transits. Our understanding of these small planets at longer periods is due to extrapolation from the period population of larger radius planets. It is possible that these small planets follow a unique distribution model, independent of their larger radius counter parts. Without a significant population of detected long period, small radius planets, extrapolations provide our best method of estimation. \

Further concern lays in the fact that the period distribution (derived by \citet{zin18}) fails to extend beyond the outer limits of the HZ in most cases (see Figure \ref{fig:hzcut}). Only 22\% of the stellar systems considered in this study have an outer HZ limit within a 500 day period (the maximum HZ extends to a period of 2782 days). This indicates that we are not entirely covering the parameter space available for habitable planets. It should be noted that the period power-law scales as $p^{-.64}$ in this region and simple extrapolation would indicate that very few planets exist at these long periods. If we simply extend our upper power-law distribution limits from 500 to 2782 day and re-run our simulation, we only find a 4\% increase in habitability with the inclusion of this longer period parameter space. However, doing so requires we assume the multiplicity remains the same. This is very unlikely and it remains difficult to estimate the expected change. Including a larger parameter space will certainly increase the number of expected planets around each star. One very rudimentary method of extrapolating multiplicity is by looking at the ratio of parameter space coverage. If we integrate Equation \ref{eq:plaw} from 0.5 to 500 days and compare that with the same equation integrated from 0.5 to 2782 days we find an increase in parameter space of 21\%. This rough extrapolation indicates we could expect a 21\% increase in planets if we extend our period range, but we caution that this correction assumes uniform multiplicity over all parameter space. We know in the solar system that period spacing becomes larger as we move outward from the sun, indicating that the rate of increased multiplicity decreases with the inclusion of a larger period parameter space. In other words, the number of expected planets will increase as we add more period parameter space, but the size of this additional multiplicity will decrease with each additional piece of parameter space added. This suggests that our calculated 21\% correction is likely much higher than the true correction. Furthermore, the outer HZ limit is set by the furthest location where the greenhouse effect can be experienced. This could be an optimistic limit, thus decreasing the HZ outer limit and reducing the need for a correction. Because of this lack of period coverage, our habitability calculations only provides a lower limit, but given the extrapolated power-law calculations, we do not expect significant changes as we extend out beyond 500 day periods.\

The power-laws derived by \citet{zin18} were also calculated under the assumption of total validation for planet candidates within the \emph{Kepler} DR25 sample. As pointed out by \citet{hsu19}, ignoring the reliability of these planets will lead to the extrapolation of an inflated $\eta_{\earth}$ values. However, this change should be small as most of the planets within this sample appear to have high validation scores \citep{tho18}. Because we expect this correction to be less significant than that caused by the lack of period space coverage, we still conclude that our $\eta_{\earth}$ values provide a lower limit.\      

 We also recognize that habitability is not limited to planets. It is possible that moons may provide the crucial ingredients for habitability. Using the fact that the gas giants of our solar system harbor several moons, \citet{hil18} argues that the existence of gas giants within the HZ of 70 \emph{Kepler} stars may provide evidence for a significant population of terrestrial moons in the HZ. Since the calculations within the current study only considers exoplanets, the discovery of a sizable number of exomoons could largely inflate the amount of habitability expected within each system. Without some understanding of the exomoon population, it is difficult to estimate how other systems may differ from our own.\

Calculating the difference between systems with at least one habitable planet and at least two habitable planet in Table \ref{tab:habTable}, the results of this study optimistically indicate $20\pm1\%$ of stars like the sun should only harbor one habitable planet. In this respect, our solar system is somewhat common. Additionally, we find that only $6.4\pm0.5\%$ of GK dwarfs will harbor more than one habitable planet and $73.9\pm0.9\%$ will not harbor any habitable planets. It is important to remember that (almost by definition) we have two planets (Mars and Venus) just beyond the parameter space of habitability. If these limits are correct, we nearly harbor a three habitable planet system, putting us among the rare $1.0\pm0.2\%$ of solar-like stars. The importance of such an architecture remains unclear and warrants further study.\

\section{Conclusions}
\label{sec:con}
Using the updated population parameters of \citet{zin18} and optimistic planetary traits required for habitability, we have provided a frequency estimate for habitable planets in the \emph{Kepler} field. We break this frequency down to account for multiplicity within the HZ (see Table \ref{tab:habTable}). Using our most optimistic radius cuts (0.72-1.7$r_{\earth}$) we find $\eta_{\earth}=0.34\pm 0.02$. This value could be larger in reality, as only 22\% of the stellar sample provide a HZ that is contained within a 500 day period, but we expect such corrections to be small.\

We find that multiplicity within the habitable parameter space should be somewhat common. Our calculation estimates that $6.4\pm0.5\%$ of solar-like stars should have more than one habitable planet with $0.036\pm0.009\%$ containing as many as five habitable planets.

This non-negligible fraction of systems expected to contain multiple habitable planets is good news for the \emph{Interplanetary Eavesdropping} program run by SETI (The Search for Extraterrestrial Intelligence).\footnote{\url{https://medium.com/ibm-watson-data-lab/trappist-1-interplanetary-eavesdropping-on-ibm-cloud-eca932561b32}} Here, as multiple habitable planet systems (such as Trappist-1) come in conjunction with the Earth, an attempt is made to obtain leaked radio transmissions between the habitable planets. Our results indicate that 6.4\% ($\sim5,500$ stars within our sample) of GK dwarfs harbor this type of architecture. It is hopeful that many more systems of multiple habitable planets should be found in the near future, providing more potential targets for this search.

\section*{Acknowledgement}
We would like to thank the anonymous referee for useful feedback. The simulations described here were performed on the UCLA Hoffman2 shared computing cluster and using the resources provided by the Bhaumik Institute. We would also like to thank the UCLA Department of Physics and Astronomy for support of this project. This research has made use of the NASA Exoplanet Archive, which is operated by the California Institute of Technology, under contract with the National Aeronautics and Space Administration under the Exoplanet Exploration Program.

\bibliographystyle{mnras}
\bibliography{paperBib}

\end{document}